%% file: main.tex
\documentclass[runningheads]{llncs}
\usepackage{ifthen,color}
\definecolor{darkgreen}{rgb} {0.0,0.5,0.0}
\definecolor{darkblue}{rgb} {0.0,0.0,0.5}
\definecolor{bluegreen}{rgb}{0.0,0.5,0.5}

\usepackage{placeins}
\usepackage{multirow}
\usepackage{microtype}
\usepackage[export]{adjustbox}
\usepackage{enumitem}
\usepackage{pdfpages}
\usepackage{verbatim}

\usepackage{graphicx}

\usepackage{amsmath}

\usepackage{cite}
\usepackage{booktabs}

\usepackage{algorithmicx,algorithm,algpseudocode}

\usepackage{url}
\begin{document}
\title{Fast Gunrock Subgraph Matching (GSM) on GPUs}

\author{Leyuan Wang\inst{1}\orcidID{0000-0002-4941-8454} \and
John D. Owens\inst{1}\orcidID{0000-0001-6582-8237}}
\authorrunning{L. Wang and J. D. Owens}
\institute{University of California, Davis CA 95616, USA}
\maketitle              % typeset the header of the contribution
\begin{abstract}
In this paper, we propose a GPU-efficient subgraph isomorphism algorithm using the Gunrock graph analytic framework, GSM (Gunrock Subgraph Matching), to compute graph matching on GPUs. In contrast to previous approaches on the CPU which are based on depth-first traversal, GSM is BFS-based: possible matches are explored simultaneously in a breadth-first strategy. The advantage of using BFS-based traversal is that we can leverage the massively parallel processing capabilities of the GPU. The disadvantage is the generation of more intermediate results. We propose several optimization techniques to cope with the problem. Our implementation follows a \emph{filtering-and-verification} strategy. While most previous work on GPUs requires one-/two-step joining, we use one-step verification to decide the candidates in current frontier of nodes. Our implementation has a speedup up to $4\times$ over previous GPU state-of-the-art implementation.

\end{abstract}
\section{Introduction}
\label{sec:intro}
\input{tex/intro}

\section{Preliminaries and Related Work}
\label{sec:related}

\input{tex/related}

\section{Approach}
\label{sec:approach}
\input{tex/approach}

\section{Experiments and Results}
\label{sec:exp}
\input{tex/exp}

\section{Conclusion}
\label{sec:conc}
\input{tex/conc}
\bibliographystyle{splncs04}
\bibliography{all, subgraph-matching}
\end{document}

%% file: tex/intro.tex
Graphs can provide meaningful representations of objects and patterns, as well as more abstract descriptions. The representative power of graphs lies in their ability to characterize multiple pieces of information, as well as the relationships between them. Because of those properties, graph data structures have been leveraged in a wide spectrum of applications including social media, the World Wide Web, biological and genetic interactions, cyber network, co-author networks, citations, etc. And at the heart of graph theory is the problem of graph matching, which attempts to find a way to map one graph onto another both topologically and semantically.
%There have been two completely different directions for supporting subgraph matching. One direction is to develop specialized query processing engines, while the other is to develop efficient subgraph isomorphism algorithms for general, labeled / unlabeled graphs. Previously, the compute power required to run these approaches have motivated their implementation on distributed CPU systems. But on a distributed system, network transfer overhead becomes a bottleneck. In order to explore the compute power of a GPU-equipped single node, we address the latter direction with our implementation of GSM.

In this work, the subgraph matching problem is defined as the enumeration of all subgraph isomorphisms between two graphs. In the general case (i.e., if no restrictive assumptions are made on the graphs), the subgraph isomorphism problem is provably NP-complete. Most previous subgraph matching algorithms on the CPU are based on depth-first backtracking which is complex and inefficient to implement on GPUs (Section~\ref{sec:related}). Existing work on on GPUs is generally bound by memory capacity.

%Many graph libraries make attempts to solve some common operations on parallel machines including the Parallel Boost Graph Library (PBGL), Pregel~\cite{Malewicz:2010:PSL}, GraphLab~\cite{Low:2010:GAN}, PowerGraph~\cite{Gonzalez:2012:PDG}, Ligra~\cite{Shun:2013:LAL}, Gunrock~\cite{Wang:2017:GGG}, and nvGraph.\footnote{nvGRAPH is available at \url{https://developer.nvidia.com/nvgraph}.} Modern graphics processors (GPUs) have been leveraged in those operations and have gained success in several parallel applications. But few of those libraries have efficiently solved the subgraph isomorphism problem, either because of the framework is not fit for the problem or because no method fully exploits the large amount of parallelism available on a GPU\@. Existing solutions are mostly based on either backtracking or filtering-and-joining mechanisms. Backtracking-based methods do not fit on GPU architectures because of their recursive nature. Existing filtering-and-joining strategies that focus on optimizing matching order either generate a large amount of intermediate results (stressing memory capacity and bandwidth) or handle intermediate results in an inefficient way.

In this work, we make the following contributions% , note that the graphs we focus on are undirected graphs with the options of having node and edge labels
:
\begin{enumerate}
\item We design a subgraph matching approach using breadth-first search as the core operation;
\item We focus on the problem of memory capacity, optimizing our algorithm to generate as few intermediate results as possible by limiting the memory usage linear to matched subgraphs; and
\item We achieve best-of-class performance on a variety of datasets and scenarios when compared to prior CPU and GPU work.
\end{enumerate}

%% file: tex/related.tex
\subsection{Prior Work}
%\john{Something that you don't note here at all is preprocessing on the data graph. We input CSR and thus aren't doing any preprocessing. But some approaches build indexes (for instance) and spend a lot of time doing that. Those indexes are not query-specific, but they are definitely preprocessing. Gunrock, philosophically, does not take a preprocessing approach. Anyway, you might note the direct competition and whether it does preprocessing or not.}
Carletti et al.~\cite{Carletti:2017:NAS} categorizes existing subgraph matching algorithms into three approaches: tree search, constraint programming, and graph indexing.

Tree search methods usually adopt a depth-first search with backtracking to formulate solultions incrementally.
%At each new iteration, a pair of nodes is added to the partial result, after checking  the addition is consistent with the constraints of subgraph isomorphism and with the specific heuristic used by the algorithm. At the point where no more pairs can be added, they backtrack by removing the last added pair and trying a new one.
One of the advantages of depth-first based methods is that the space it takes to store intermediate results is proportional to the number of nodes and thus these algorithms have a linear memory complexity. But this method is difficult to parallelize because it is recursive and needs extra scheduling in order to make use of many parallel cores; this is not a GPU-friendly approach. Previous works that use this method include the earliest subgraph isomorphism solution by Ullmann~\cite{Ullmann:1976:ASI}  as well as later optimizations including VF2~\cite{P.Cordella:2004:GIA}, QuickSI~\cite{Shang:2008:TVH}, VF2Plus~\cite{Carletti:2015:VF2+}, VF3~\cite{Carletti:2017:NAS} and VF3P~\cite{Carletti:2019:VF3P}.

The second category of algorithms, based on constraint programming, see the search for subgraph isomorphism as a constraint satisfaction problem, with the goal to find an assignment of values to a set of variables that satisfies a set of mutual constraints.
%In particular, for each node of the pattern a domain of compatibility is maintained, containing the potential matching nodes in the target. Local constraints (e.g., node or edge consistency) are propagated to different parts of the graphs to reduce the domains, until only few candidate matches remain that can be explored to find the solutions.
An early algorithm following this approach is McGregor's~\cite{MCGREGOR:1979:RCA}; more recent proposals are proposed by Zampelli et al.~\cite{Zampelli:2010:SSI}, Solnon et al.~\cite{Solnon:2010:AFS}, and Ullmann~\cite{Ullmann:2011:BAB}.

The last category, graph indexing, originates from graph database applications, where the goal is to retrieve, from a large set of graphs, only the ones containing the desired pattern. To achieve this aim, an index structure is built which makes it possible to quickly verify if the pattern is present or not in a target graph, usually without even requiring to load the whole target in memory, and thus filtering out unfruitful targets. In general, after index verification is passed, a more costly refinement phase is needed to actually determine whether and where the pattern graph is present. GADDI~\cite{Zhang:2009:GDI}, GraphQL~\cite{He:2008:GQL}, SPath~\cite{Zhao:2010:GQO}, and Turbo$_{\text{ISO}}$~\cite{Han:2013:TTU} are recent implementations based on this approach.

GPU subgraph matching efforts have primarily focused on the latter two categories of algorithms but face the same limitation: generating too many intermediate results and thus being limited by the GPU's modest memory capacity. Also, graph-indexing methods spend a lot of time building an index to the data graph as a preprocess. While preprocessing methods are cost-effective for repeated queries into the same graph, they are expensive for a more limited number of queries. In contrast, our approach inputs data graphs in CSR format and does no preprocessing on the data graph.

\subsubsection{GPU-based Subgraph Matching}
\label{sec:gpu}
Recently, GPUs with massively parallel processing architectures have been successfully leveraged for fundamental graph operations on large graphs.  Subgraph matching implementations using GPUs include GPU-STwig~\cite{Lin:2014:ESM}, GpSM~\cite{Tran:2015:FSM,Tran:2016:TGC}, and GSI~\cite{Zeng:2019:GSI}. Traditional backtracking approaches for subgraph matching cannot be efficiently adapted to GPUs mainly due to two problems. First, GPU operations are based on warps---groups of threads executed in single-instruction-multiple-data (SIMD) fashion---and different execution paths generated by backtracking algorithms can cause warp divergence. Second, GPU implementations for coalesced memory accesses are no longer straightforward due to irregular access patterns. The aforementioned four methods address these difficulties and propose numerous optimizations on GPUs.

%One of the earliest implementations on GPUs is from Lin et al.~\cite{Lin:2014:ESM}, based on the distributed CPU STwig algorithm. They make use of elaborative joining order selection and a fully pipelined mechanism. They analyze the bottleneck of joining on GPUs and point out the problem of the large intermediate data produced in a multi-way join. So in the paper, they only focus on joining STwig on GPUs, where they make use of a hash table to improve the efficiency.
%They put the decomposition and STwig matching steps on the CPU and pipeline those steps with the joining steps on the GPU\@.
%However, the complexity of the algorithm, compared with the original STwig on the CPU, is not reduced, and  memory transfers between CPU and GPU increase their steps. Though 2-way joins produce fewer redundant results, the consumption of GPU memory is still prohibitively large.
GpSM~\cite{Tran:2015:FSM,Tran:2016:TGC} follows a \emph{filtering-and-joining} strategy.  Their filtering step is similar to Turbo$_{\text{ISO}}$,  ranking  query nodes based on the possible number of candidate nodes, and  using neighborhood exploration to further prune out invalid candidates. But for some datasets whose candidate nodes cannot simply be pruned out based on node labels and degree information, the joining step will still be a bottleneck, which turns their algorithm into a memory-bound one. In Section~\ref{sec:exp}, we show that our GSM has better scalability than GpSM and can process larger graphs than GpSM's upper limit.

A simpler but similar task to subgraph matching is triangle counting, which enumerates the triangles in a graph. Recent work from Wang et al.~\cite{Wang:2016:ACS,Wang:2016:FSM} in this area uses a more general subgraph matching approach. It is compared with two other algorithms that only support querying triangles and are specially optimized for that purpose. Wang et al.'s implementation is also based on \emph{filtering-and-joining},  decomposing the query to edges and joining the corresponding candidate edges to compose final subgraphs. This implementation successfully exploits GPU parallelism by distributing the work of finding candidate edges for query graphs across all GPU processors. But its bottleneck is joining, which generates an exponential number of intermediate results and quickly runs into a GPU memory-capacity limitation. And performance-wise, this implementation compares poorly to triangle-matching-specific algorithms. Their recent update to their implementation~\cite{Wang:2019:FBT} for the triangle-counting GraphChallenge outperforms the previous year's champion; our work here extends their approach to generalized subgraphs.

GSI is a recent work (with publication slated for April 2020) on GPUs from Zeng et al.~\cite{Zeng:2019:GSI}. This work follows a \emph{filtering-and-joining} strategy and describes several useful optimizations. One cornerstone of their effort is a different data structure for graph storage, PCSR, which optimizes matching latency at the cost of preprocessing the input data graph. Note that this preprocessing time is not added to the final matching time shown in their paper (and also is not counted in our Section~\ref{sec:exp}). In contrast, our implementation has no such preprocessing cost and thus our runtimes in Section~\ref{sec:exp} account for all processing. In our analysis, we found GSI is currently the fastest subgraph matching method on the GPU\@. However, it performs poorly on unlabeled graphs, running out of memory even on small datasets like \emph{Enron} (69k vertices and 549k edges) if the graph is not labeled. The reason is their inefficient storage of intermediate results. We compare with their performance in Section~\ref{sec:exp} and show that our GSM has better performance and scalability on both labeled and unlabeled graphs.

%% file: tex/approach.tex
Our algorithm is an extension of the triangle counting work from Wang et al.~\cite{Wang:2019:FBT} which follows a \emph{filtering-and-verification} strategy. The \emph{filtering} process prunes out candidates which cannot contribute to final matches based on certain constrains such as degree, label, and connections. We will describe why optimization in this process can lead to more efficiency in both running time as well as memory usage in this section. The \emph{verification} process on previous CPU-based algorithms searches matching between the query graph and the filtered data graph in a depth-first manner which follows Ullman's~\cite{Ullmann:2011:BAB} backtracking method. For GPU-based methods, the \emph{verification} process proceeds in a breath-first manner to do simultaneous constraints verification on massive parallel processors. That is also what we leverage in this work. Compared with the depth-first backtracking subroutine, BFS-based verification has the disadvantage of generating more intermediate results and doing more redundant traversals. We will illustrate how we address those problems in our implementation.

%Most existing subgraph matching approaches follow either a \emph{filtering-and-verification} or \emph{filtering-and-joining} strategy. The filtering step prunes out candidates that cannot contribute to the final solutions; we later show why this step determines the efficiency of the algorithm. The verification step is generally based on Ullman's backtracking subroutine, which searches in a depth-first manner for matching between the query graph and the updated data graph obtained from the filtering step. The joining step combines the filtered candidate edges or partial results to the complete matched subgraphs by checking certain constraints. Previous graph-index-based methods mostly follow \emph{filtering-and-joining}, which was also our approach when we initially designed our algorithm. But we found that the large amount of intermediate results generated during the \emph{joining} step are a challenge to store on a single GPU\@. Instead, the optimized method that we propose in this paper follows a \emph{filtering-and-verification} approach, but unlike previous works, it is not based on depth-first search.

An efficient \emph{filtering} process can save a lot of effort by reducing the search space for later \emph{verification} process. Especially for BFS-based methods, the search space could be very large without pruning out non-fruitful nodes and edges which makes the \emph{verification} process become a bottleneck for the whole implementation. To cope with the problem, we use an optimization method called neighborhood encoding in our \emph{filtering} process. The encoding is updated after each local refinement. And the refinement process is repeated for several times which is set as a parameter in our implementation. In this way, more invalid candidates are pruned and a more effective global search space is generated. Detailed implementation is illustrated in section~\ref{subsec:algorithm}.
%The aim of the filtering step is to reduce the search space on which later verification or joining steps operate. A good filtering technique can save significant effort by pruning out non-valid nodes/edges before verification or joining, which is usually the bottleneck of the whole algorithm. Filtering mechanisms can be classified into two categories depending on their exploring scopes: local or global. A local refinement mechanism prunes the set of mappings that are candidates for each single vertex. A global pruning reduces the global search space. In our approach, we use an effective filtering method by footprint encoding that is updated after each local pruning and thus generates a more effective global pruning of the search space.
We also reduce intermediate results by using an elaborate query order selection, which is borrowed from previous works. A novel idea we propose in this paper on the selection of query node visiting sequence is that we maintain a set of constraints on node ID values of the query graph in order to avoid generating partial results, which eventually become duplicated combinations of nodes of the same subgraph.
%For instance, if we want to query a triangle in the data graph, and we have one subgraph match represented as $[1,2,3]$, without our method, the final results can contain up to six enumerations (in different order) of the same three nodes.

For the \emph{verification} process, we do breath-first traversals starting from every node in the data graph in parallel (all-source-BFS). In each iteration of BFS traversal, we verify whether the newly visited nodes in data graph satisfy connection constrains with previously visited nodes in partial results as well as the constrains defined by the query graph. Note that in BFS, the nodes visited in each iteration increases in an exponential scale. In order to avoid the huge expansion of memory with the increase of iterations, we use two techniques. First, do a compaction after each iteration to prune out non-valid partial results. Second, we store the partial results in a compressed way. Details are illustrated in section~\ref{subsec:algorithm}.

While we believe that our method is not specific to any particular graph framework, we leverage the Gunrock framework with our work and describe it first.

\subsection{Gunrock graph processing framework}
The Gunrock~\cite{Wang:2017:GGG} GPU-based graph analytics framework uses a high-level, bulk-synchronous, data-centric abstraction. Gunrock programs are expressed as manipulations of frontiers of vertices or edges that are actively participating in the computation. Its operators currently include: \emph{advance}, which generates a new frontier by visiting the neighboring vertices/edges to the current frontier (applying work distribution/load balancing techniques for efficiency); \emph{filter}, which removes elements from a frontier via validation tests; \emph{segmented intersection}, which computes the intersection of two input frontiers; and
\emph{compute}, which computes user-defined vertex/edge-centric functions that run in parallel over elements in the frontier; these functions can be fused with advance or filter operators.
%\begin{description}
%  \item[Advance] which generates a new frontier by visiting the neighboring vertices/edges to the current frontier (applying work distribution/load balancing techniques for efficiency).
%  \item[Filter] which removes elements from a frontier via validation tests.
%  \item[Segmented intersection] which computes the intersection of two input frontiers.
%  \item[Compute] which computes user-defined vertex/edge-centric functions that run in parallel over elements in the frontier; these functions can be fused with advance or filter operators.
%\end{description}

%The Gunrock framework is very efficient for BFS-based algorithms. Since our algorithm is BFS-based, we are able to fully utilize the massive parallelism of GPUs and construct our approach using the above four operators.
%We also leverage Gunrock's capability to support frontiers of either nodes or edges.
By using the above operators, we are able to leverage Gunrock's efficiency on bread-first traversal in our implementation. We elaborate on this design choice in our approach described in the following section.

\subsection{Our proposed algorithm: GSM}
\label{subsec:algorithm}
%The algorithm we propose achieves the following goals:
%\begin{enumerate}
%\item Scalable on GPU cores.
%\item Linear memory complexity to the number of matched subgraphs.
%\end{enumerate}

%Unlike most of the previous methods that are based on either tree search or graph indexing, our algorithm is BFS-based, which provides us the opportunity to efficiently leverage the massive parallelism of the GPU\@. In terms of memory usage, though the algorithm is not recursive, we are still able to limit the space needed proportional to the number of edges in the graph by using more efficient pruning techniques, by selecting query order in a novel way and by efficient compaction method and hash table values for storing intermediate results.
Our GSM is scalable with the number GPU cores and consumes linear memory proportional to the number of matched subgraphs. Algorithm~\ref{alg:sm} is the pseudocode of GSM, which also shows the implementation using the Gunrock framework. There two inputs for our implementation: a (small) query graph $Q$ and a (large) data graph $G$ for searching. The input graphs are undirected and they could be either labeled (with node/edge labels) or unlabeled. There are two outputs: the subgraph counting and subgraph enumeration.

\begin{algorithm}[!ht] \caption{GSM}
\label{alg:sm}
\renewcommand{\algorithmicrequire}{\textbf{Input:}}
\renewcommand{\algorithmicensure}{\textbf{Output:}}
        \begin{small}
                \begin{algorithmic}[1]
                  \Require{Query Graph $Q$, Data Graph $G$.}
           \Ensure{Number of isomorphic subgraphs $n$ and listings of all subgraphs.}
           \Procedure{PreCompute\_on\_CPUs}{}
                                \State
                                 \Call{Compute\_query\_node\_sequence\_info}{$Q, d_M, P_f, \mathit{deg}$}
                                 \State
                                 \Call{Store\_node\_neighborhood\_connection}{$\mathit{nn}$}
                                 \State
                                 \Call{Store\_none\_tree\_edges}{$\mathit{ne}$}
                      %           \State
                                 %\Call{Generate\_UMO}{$\mathit{ns}$}
                                 \EndProcedure
                        \Procedure{Filter\_candidate\_set}{$Q,G$}
                        \State
                        \Call{Advance+Compute}{$G$}\Comment{Compute NE for each node in $G$.}
                        \State
                        \Call{Filter+Compute}{$G,Q,\mathit{c\_set}$}\Comment{Filter nodes based on $Q$'s (NE, label, $\mathit{deg}$); update $G$'s (NE, $\mathit{deg}$); and write to candidate set $\mathit{c\_set}$.}
                        \EndProcedure
                        \While{$(|M[i]| < |Q|)$} \Comment{The number of verify iterations is the size of $Q$}
                        \Procedure{Verify\_Constraints}{$G,Q,\mathit{c\_set}, M$}
                        \State
                        \Call{Advance}{$\mathit{c\_set}$}\Comment{All-source BFS traversal from $\mathit{c\_set}$ to dest(neighbor) nodes which are verified on stored constraints.}
                        \State
                        \Call{Compute}{$\mathit{c\_set}$}\Comment{Compact satisfied dest nodes to $\mathit{c\_set}$.}
                        \State
                        \Call{Write\_to\_Partial}{$M$}\Comment{Combine $\mathit{c\_set}$ with partial results $M$.}
                        \EndProcedure
                        \EndWhile
                        \State
                        \Return{Number of subgraphs :$\frac{|M|}{|Q|}$, and subgraph enumeration $M$}
                \end{algorithmic}
        \end{small}
\end{algorithm}

First we preprocess the query nodes' order selection on the CPU\@. This preprocessing cost is much cheaper than prior work's preprocessing on entire data graphs.  Basically, we want to generate a node visiting order, a permutation of nodes in $Q$, in order to give priority to nodes having more constraints, such as having smaller probabilities of finding a match in graph $G$ or having more connections to already matched nodes in $Q$, so that those constraints can be applied in earlier stages to reduce the global search space in $G$. We borrow ideas from VF3~\cite{Carletti:2017:NAS} to precompute the order based on the Maximum Likelihood Estimation (MLE) method. The MLE of finding a match for a node $u$ in $Q$ to a node $v$ in $G$ is determined by label frequency and node degree. They estimate for each node $u\in Q$, the probability $P_f(u)$ to find a node $v\in G$ that is compatible with $u$, meaning sharing the same label with $u$ and have a degree equal to or larger than $u$'s degree. Please refer to the original paper for a detailed definition of $P_f$.
%expressed as follows.
%\begin{equation}
%P_f(u) = Pr(l'(v) == l(u), d(v)\geq d(u))
%\end{equation}
%Note that this is only a weak estimator to be computed with low temporal and spatial complexity since the accuracy only affects the exploration order of the query nodes instead of the final results. And since the label frequency and degree constraint are independent,  they can be computed separately, which makes the equation's complexity linear shown in the following equation.
%\begin{equation}
%P_f(u) = P_l(l(u)) \cdot \Sigma_{d'(v)\geq %d(u)}P_d(d')
%\end{equation}
The structure constraints brought by nodes already mapped can be addressed in node mapping degrees $d_M$ (referred in line 2 of Alg.~\ref{alg:sm}), which is a new concept proposed in VF3\@. $d_M$ is defined as the number of edges connected to the nodes already in the partial result. Note that at each step when a node is added to the partial result, $d_M$  needs to be updated for all remaining nodes that do not belong to the partial result. So the query order is determined in the following priority order: $d_M$, $P_f$, and $\mathit{deg}$ (shown in Alg.~\ref{alg:sm}). If all three of the evaluations are equal, the order is chosen arbitrarily. The query node sequence we generate is based on bread-first traveral of a spanning tree.
%Besides precomputing the query node sequence, we also store the query graph connection information as part of our stored constrains. 
%We make sure that the node sequence we generate meets the BFS traversal of a spanning derived from $Q$\@. So 
We store each nodes' visited parents in the spanning tree (line 3 of Alg.~\ref{alg:sm}) as well as any non-tree edge connection (line 4 of Alg.~\ref{alg:sm}).

The main algorithm contains two processes: \emph{filtering} (line 6 of Alg.~\ref{alg:sm}) and \emph{verification} (line 11 of Alg.~\ref{alg:sm}). The \emph{filtering} process starts with a computation of neighborhood encoding (NE) (line 7 of Alg.~\ref{alg:sm}), which is computed based on both the degrees and labels of neighboring nodes in the data graph. The definition of NE is the sum of the labels of the nodes' neighbors. If node labels are not defined in a given use case, NE will become the degree of $u$, assuming each node's label is one. NE is computed for each node in both the query graph and the data graph. The computation of NE of the small query graph is done on the CPU during preprocssing. And we compute NE for the large data graph on the GPU during \emph{filtering} process. The purpose of NE is to maintain neighborhood information for each node in a compressed way. And based on NE, degree and label information, we can prune out invalid node candidates (line 8 of Alg.~\ref{alg:sm}). Note that after each iteration of pruning, NE and degree information of the data graph is updated. The number of pruning is a parameter which can be set in our implementation.
%NE is based on an idea of using a value to represent neighborhood information that characterizes each vertex in the data graph. We compute each query node's NE during preprocessing on the CPU and filter the candidate nodes in $G$ based on NE, label, and degree (line 8 of Alg.~\ref{alg:sm}). Note that both NE and degree information are updated once we filter out non-valid candidate nodes.

During the \emph{verification} process, we do multi-source breadth-first traversal. The number of traversals equals the number of query nodes (line 10 of Alg.~\ref{alg:sm}). The source nodes are from the candidate set ($\mathit{c\_set}$) which is the result of the previous \emph{filtering} process.  In each traversal, we verify whether the destination nodes are valid based on stored constraints including connections with existing nodes in partial results as well as non-tree connections from the query graph (line 12 of Alg.~\ref{alg:sm}). We avoid generating excessive intermediate results by two optimizations. First, we do a compaction of the updated $\mathit{c\_set}$ before writing them to partial results (line 13 of Alg.~\ref{alg:sm}). Second, we store the listings of matched candidates in each iteration in terms of values instead of node ID tuples. We use a hash function to convert the node ID combinations to a certain value, which reduces the memory usage for storing intermediate results. Note that after each iteration, the partial results with size less than iteration number will be automatically filtered out and won't be passed to next iteration.

\subsection{Implementation}
\begin{figure}[ht]
\begin{center}
\includegraphics[width=0.8\columnwidth]{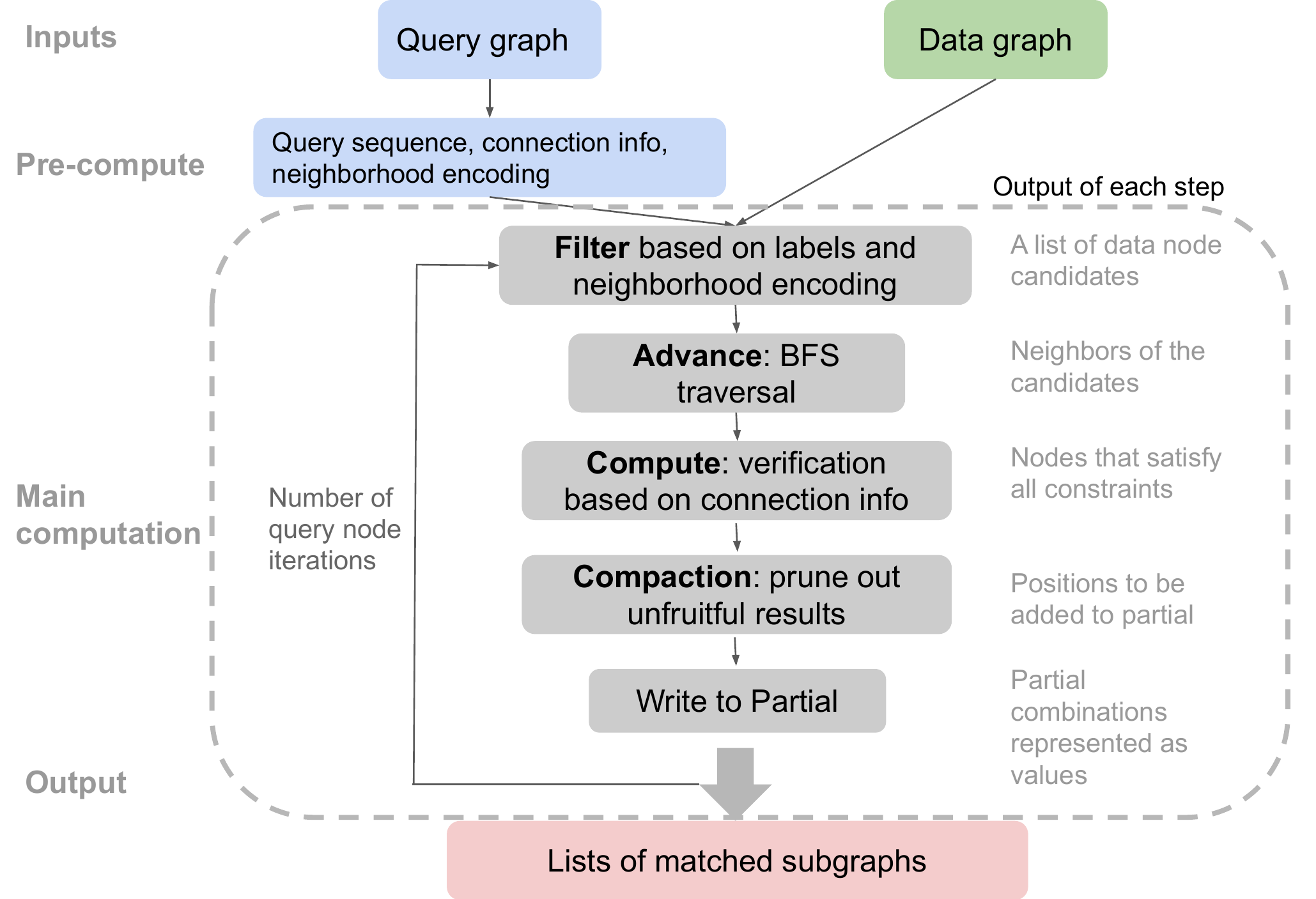}
\caption{GSM implementation flow chart. \label{fig:sm}}
\end{center}
\end{figure}

Figure~\ref{fig:sm} further explains the workflow of Alg.~\ref{alg:sm} in an operator-level.
We store graphs
%(both query graph and data graph)
in a space-efficient fashion on the GPU by using compressed sparse row (CSR)\@.
%We require no preprocessing on this graph, making it suitable for direct input from other stages in a data-analytics pipeline.
Three operators from the Gunrock framewore are used throughout our implementation: filter, advance and compute.
During main computation, a filter operator is called first to prune data graph nodes based on precomputed query node's information. The result is a candidate set. Then an advance operator is used to traverse the neighbors of nodes in the candidate set. Next, we use a compute operator to verify if the newly visited nodes satisfy connection constrains. Note that the advance operator from Gunrock maps the newly traversed edges to consecutive GPU threads and thus the compute operator is executed in a SIMT manner. The result is an updated candidate set containing the adjacent nodes of previous candidate set which pass the verification process. Then a compaction operation is used to select the new candidate nodes from scattered GPU threads to consecutive positions in order to serve as the inputs for the ease of the next iteration of advance traversal.
%First, the filter operator inputs all nodes from the data graph $G$ and returns nodes with satisfied NE, label, and degree requirements according to the precomputed query graph $Q$'s information. Next, we use the advance operator to traverse all the neighbor nodes of the candidate nodes resulting from the previous step. Note that we traverse the query graph and the data graph at the same time. So we must verify if the newly traversed edges and the data graph satisfy connection constraints with their corresponding query edges. Gunrock's advance operator allows us to compute this traversal in a massively parallel manner, where the large number of newly added edges is mapped to consecutive GPU threads and the verification computation is also done in a SIMT manner. The output is neighbor nodes from the threads that pass the constraint-verification tests of the advance operator. In the next step, we run a compaction on the candidate nodes from scattered threads to consecutive positions in order to serve as the inputs for the ease of the next iteration of advance traversal.
When writing to partial results, instead of directly writing the node ID information to intermediate results, we use a hash function to map the current node combinations to a value, and we store the value to partial results to further save memory usage.
Note that in GPU computing, consecutive reads perform much better than scattered reads. So not only do we benefit from better memory complexity but also gain a memory performance boost from compaction and hashing. We show in Section~\ref{sec:exp} that our implementation is currently the most memory-efficient GPU implementation.

We use $k$-look-ahead as an optimization method in the implementation of the \emph{verification} process. The idea is borrowed from the VF3 algorithm. The purpose of this mechanism is to prune out more redundant partial results in each iteration based on feasibility rules besides constraint verification. $k$-look-ahead is proposed based on a concept that a non-consistent state will not generate any successive consistent states. But in contrast, even if a state is consistent in current step, it may not generate any consistent descendants after a number of steps, and thus cannot contribute any fruitful results. The detection of such situations can help us further prune out more invalid intermediate results. A $k$-look-ahead is thus defined as a mechanism to detect a state which won't have any consistent descendants $k$ step ahead.
%The foundation of $k$-look-ahead is that it is possible to prove a non-consistent state will not generate any successive consistent states. However, it may happen that, even if the state is consistent according to our constraints verification, after a certain number of steps it cannot generate any consistent descendants and, thus, cannot contribute to the final results. The detection of such situations would help to further reduce the number of explored states by avoiding the generation of consistent, but unfruitful, states. In order to achieve this goal, the algorithm verifies if the addition of the new candidate nodes to the partial results generates a consistent state; in addition, it is able to detect $k$ steps in advance if the state will not have any consistent descendants, a $k$-look-ahead.
Note that this optimization is a necessary but not sufficient condition. If it is false, it guarantees that the current candidate node will not pass the next iteration of verification. In our implementation, we use 1- and 2-look-ahead only, which we find best balances the benefit of pruning against the cost of look-ahead.

%Another optimization is our method of avoiding duplicated final results. The hard case is when we generate a match through two different paths; we may not be able to detect the duplication through partial results alone. If this is the case, we incur both additional computation and storage during the computation. One possible solution is to use a hash table to keep track of all node sequences. Though it stops the generation of duplicate results, it cannot filter out unfruitful intermediate results. Moreover, such a large hash table storing all partial solutions would be very hard to fit on a single GPU\@. Instead we add constraints to the query node visiting order to ensure that no duplicate final results would be generated from the beginning. We successfully transfer the idea of the node equivalence class (NEC), which we previously used to reduce the depth-first search space, to solve this problem (mentioned in Section~\ref{subsec:algorithm}). For example, consider a triangle: all three nodes of a triangle are equivalent. So we need to define a visiting order of those equivalent nodes to avoid visiting the same combination multiple times. So in this example, we relate the visiting order to node id value and only visit in the order of increasing node id values. In this way, the same combination of nodes will only be visited once in the data graph.

To summarize, we use the following optimizations within our implementation of GSM:
\begin{enumerate}
\item Using neighborhood encoding to represent neighborhood information for more efficient \emph{filtering} process;
\item Using $k$-look-ahead (1-/2-look-ahead) in the \emph{verification} process to prune out more unfruitful intermediate results;
\item  Compaction of partial results after each advance operation to be stored consecutively memory, which improves memory usage and data access efficiency for the next iteration; and
\item Storing values derived from a hash function for partial results instead of storing listings of node ID tuples to further improve memory efficiency.
%\item Using node equivalence to avoid generating duplicated solutions and redundant intermediate results.
\end{enumerate}

%% file: tex/exp.tex
% experiments
To evaluate the performance of GSM, we performed experiments on both real-world datasets as well as synthetic datasets. We compare with previous state-of-the-art work, including subgraph matching on the CPU (VF3P~\cite{Carletti:2019:VF3P}) and implementations that target GPUs (GpSM by Tran et al.~\cite{Tran:2015:FSM}, Wang et al.~\cite{Wang:2016:FSM}  and GSI~\cite{Zeng:2019:GSI} by Zeng et al.). We validate the correctness of our results by comparing with Boost v1.72.0's VF2 implementation on the CPU\@.\footnote{Boost's output does not filter out duplicated combinations of matched subgraphs. In other words, they output the same combination of nodes in different orders and count them as different outputs. We postprocess Boost's results with a hash map that filters out duplicate results in order to get the correct subgraph count. Also, we found that Boost's VF2 implementation can give incorrect results when the base graph contains self-loops.}
\subsection{Datasets}
We tested our implementation on both synthetic and real-world datasets. The real-world datasets include the Enron email communication network (\emph{enron}), the Gowalla location-based social network (\emph{gowalla}), a patent citation network (\emph{patent}), and the road\_central USA road network (\emph{road\_central}). All are available from the Stanford Large Network Dataset Collection.\footnote{\url{https://snap.stanford.edu/data/}} The properties of the graphs are listed in Table~\ref{tab:dataset}.
\begin{table}
  \caption{Real-world Dataset Description Table. The edge number shown is the number of directed edges when the graphs are treated as undirected graphs and de-duplicate the redundant edges. Graph types are: r: real-world, s: scale-free, and m: mesh-like.\label{tab:dataset}}
  \centering
  \setlength{\tabcolsep}{3pt}
  \begin{tabular}{lccccc} \toprule Dataset &Vertices&Edges&Triangles&Max Degree &Type\\
    \midrule
    enron & 69,244 & 276,143 &1,067,993 & 1394 & rs
    \\ gowalla & 196,578 & 1,900,654 &  2,273,138& 14,730 & rs
    \\ patent & 3,774,768 & 16,518,948&7,515,022 & 793 & rs
    \\ road\_central & 14,081,816 & 33,866,826&228,918 & 8 & rm
    \\ \bottomrule
  \end{tabular}

\end{table}

The synthetic datasets we used are from the 10th DIMACS Graph Challenge\footnote{\url{https://sparse.tamu.edu/DIMACS10}} in the SuiteSparse Matrix Collection from the Graph500 Challenge. The sets we used are called \emph{delaunay} and \emph{kronecker}. The size of the \emph{delaunay} graph ranges from $2^{10}$ vertices (\emph{delaunay\_n10}) to $2^{24}$ vertices (\emph{delaunay\_n24}). The size of the \emph{kronecker} graphs range from 65k vertices and 2.4M edges to 2.1M vertices and 91M edges (\emph{kron\_g500-logn16} to \emph{kron\_g500-logn21}).
\subsection{Environment}
We performed all of the GPU tests on an NVIDIA TitanV GPU with 12~GB HBM2 memory capacity and 652.8~GB/s memory bandwidth. The CPU on this machine was an Intel Xeon CPU E5-2637v2 @ 3.50~GHz. GSI's current source code appears to compile and run successfully only on TitanV and TitanXP GPUs, so our comparisons to GSI use TitanV\@.

\subsection{Performance results and analysis}
\begin{figure}[ht]
  \centering
  \includegraphics[width=\linewidth]{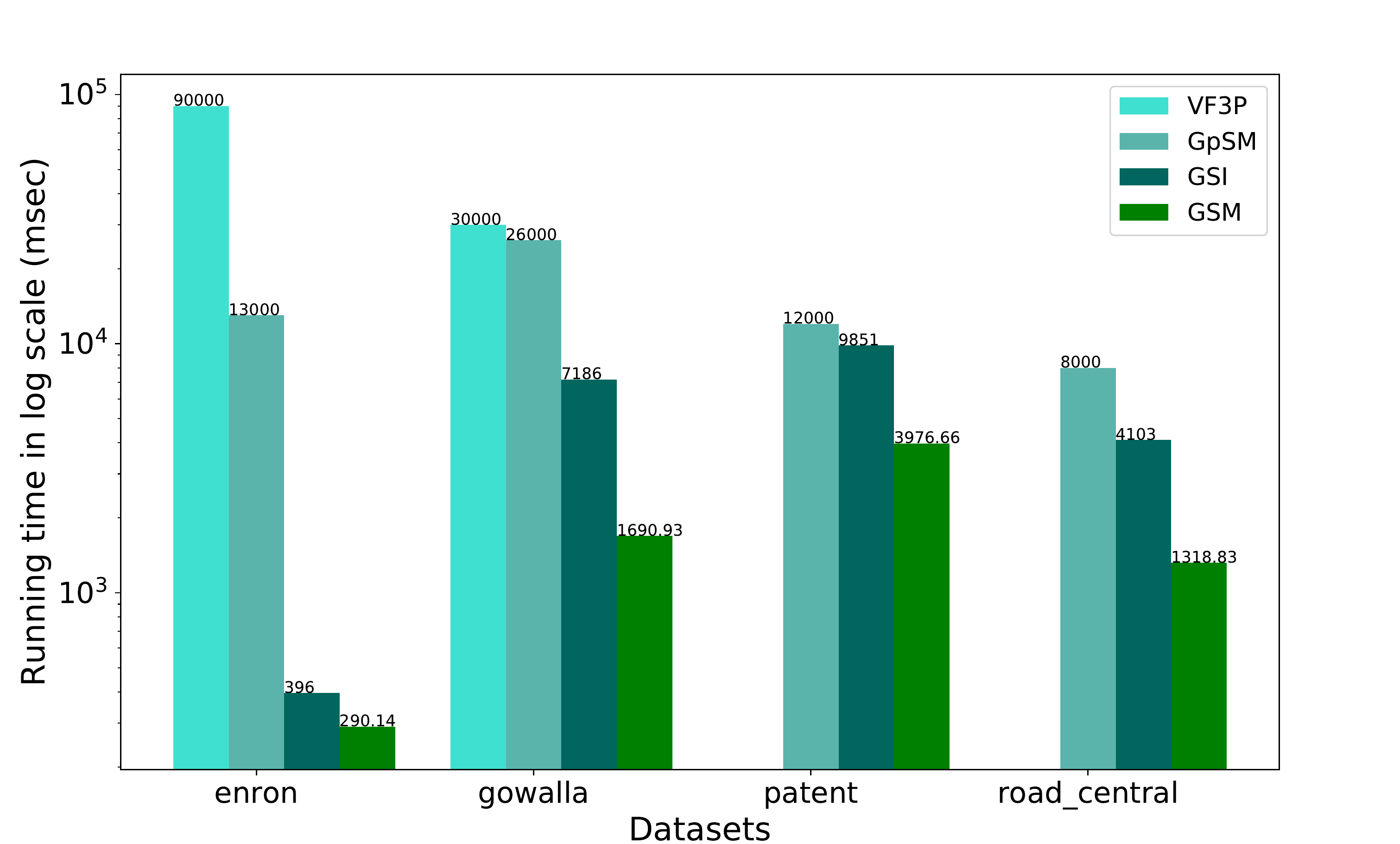}
  \caption{Runtime comparisons with previous work matching triangles on an unlabeled dataset. VF3P took around 3 hours to finish on \emph{patent} and \emph{road\_central} network; those two measurements are not included on the above graph. Dataset sizes are limited to those that GSI can run successfully without running out of memory.}
  \label{fig:perf_speedup}
\end{figure}

\paragraph{Performance with a triangle query graph}
We compared our work with previous state-of-the-art GPU counterparts (GpSM~\cite{Tran:2016:TGC} and GSI~\cite{Zeng:2019:GSI}) as well as the CPU implementation VF3P~\cite{Carletti:2019:VF3P}, summarized in Figure~\ref{fig:perf_speedup}. In this experiment, we used a triangle as our query graph, without labels on either the query graph or the data graph.\footnote{GSI runs out of memory on larger query graphs.} V3P is the slowest on every dataset, and GpSM is the slowest GPU implementation. Our GSM work is consistently faster than each of these implementations. Note that  \emph{road\_central} is $10\times$ larger than \emph{gowalla}  but its runtime is smaller. Our pruning strategies are particularly effective at pruning work from road\_central, which has fewer matches, and the join phase has less work for \emph{road\_central} than \emph{gowalla}. In general, runtime is correlated with both graph size and the number of matched subgraphs.

\paragraph{Scalability of query-graph size and label count for label-free and labeled graphs}
For our scalability tests with increasing graph sizes, we first focus on workloads without labels. Because labels reduce the number of potential matches, labelless workloads stress scalability the most, both in terms of computation time as well as memory footprint. We specifically address demanding workloads such as these with our pruning and compression optimizations. Also, we show scalability with an increasing number of labels, which highlights the performance of our label pruning method. Next, we show the results of our scalability tests.

\begin{figure}
  \centering
  \begin{minipage}{0.43\columnwidth}
    \includegraphics[width=\linewidth]{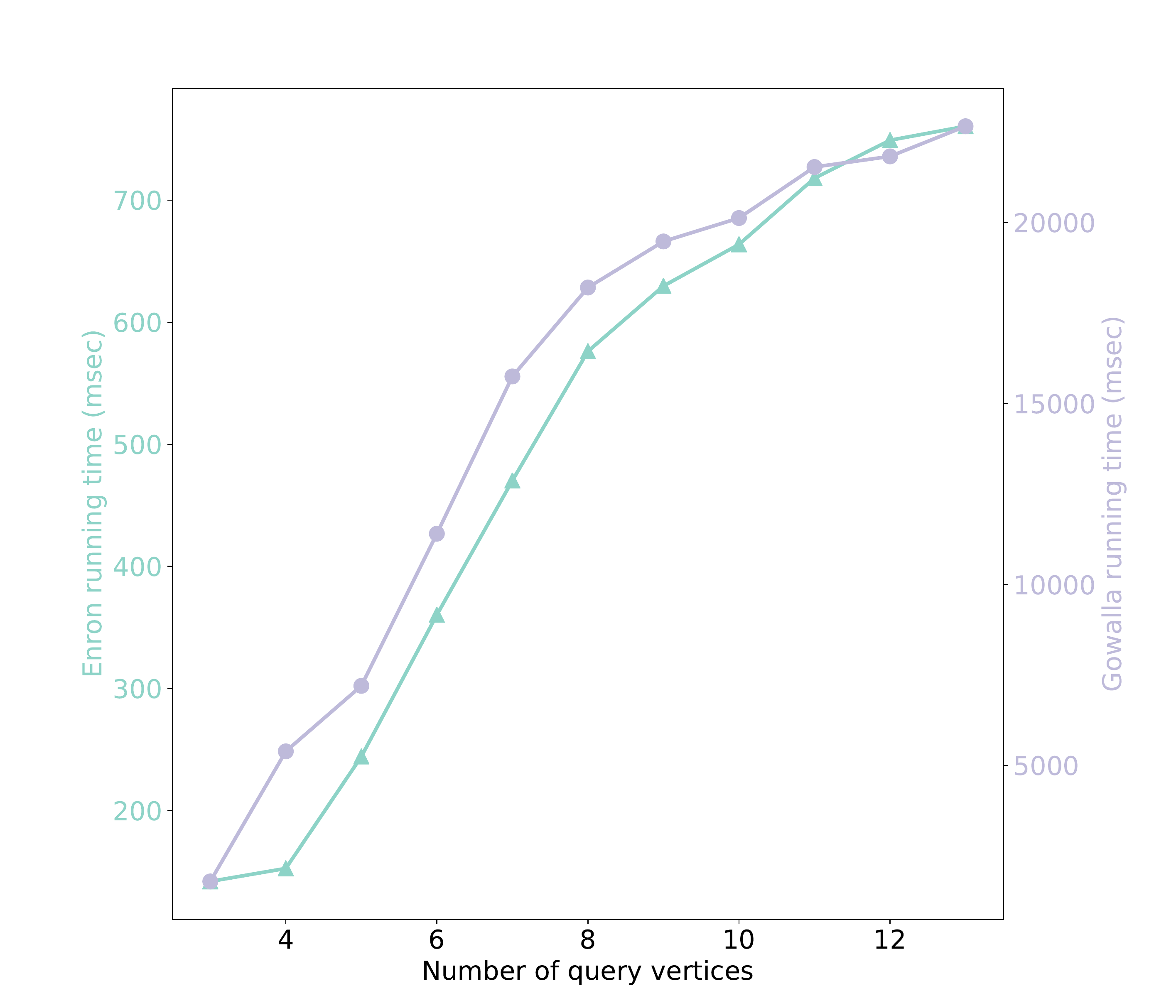}
  \end{minipage}
  \begin{minipage}{0.43\columnwidth}
    \includegraphics[width=\linewidth]{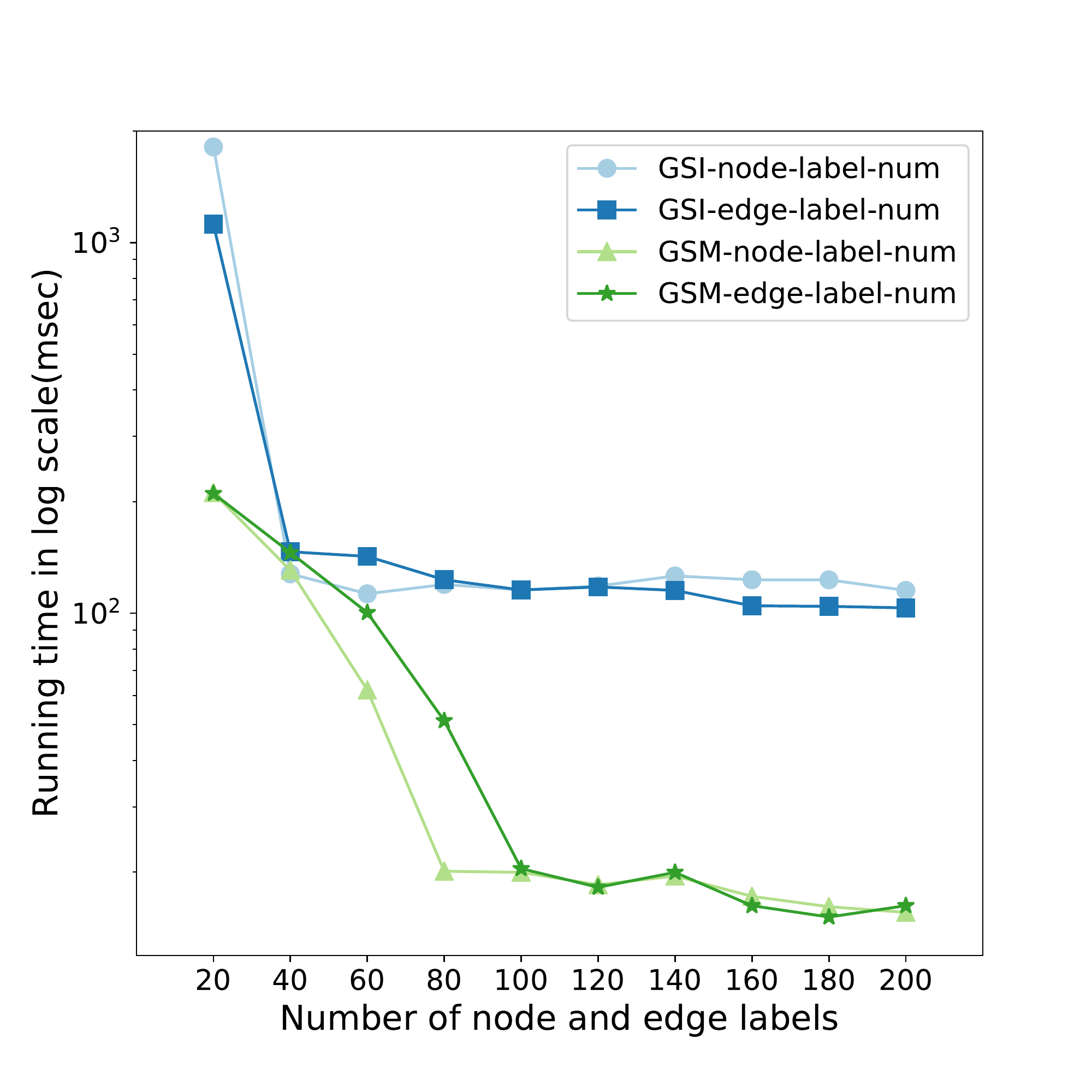}
  \end{minipage}
  \caption{Left: Scalability test of GSM on increasing query graph Enron and Gowalla, size from 3 vertices to 13 vertices. We confirmed with GSI's authors that GSI has memory-access or out-of-memory errors for any query sizes larger than 4, so we do not compare to it here. Right: Scalability test of GSM and GSI on increasing number of node and edge labels from 20 to 200 on Gowalla dataset.
  }
  \label{fig:scale-real}
\end{figure}

To explore query-graph size scalability, we choose two real-world datasets (Enron and Gowalla).  network for query node size scalability test (Figure~\ref{fig:scale-real}, left). In general, larger query sizes result in fewer matches, fewer intermediate results, and less verification, so while larger queries are more complex, the reduction in intermediate and final results yields a more modest growth in runtime.

For label-size scalability tests, we choose the Gowalla network as the data graph and use the scalability experiment of Yan et al.~\cite[Section~6.2]{Yan:2004:GIF} work (a test also used in subsequent work~\cite{Zeng:2019:GSI,Han:2010:IFC}). Briefly, this test seeds the data graph with power-law-distributed node and edge labels, then uses random walks in the data graph to create a query graph of a specified size. For each test, we generate 10 different queries with 12 nodes and 22 edges. We ran each query 10 times and used the mean runtime as the result (Figure~\ref{fig:scale-real}, right). From the figure, we can see that our GSM's runtime linearly decreases with label count from 20--100 labels, then levels off as the size of the candidate sets is not large enough to be fully parallelized. GSM demonstrates significantly better performance than GSI on this test.

\begin{figure}
  \centering
  \begin{minipage}{0.43\columnwidth}
    \includegraphics[width=\linewidth]{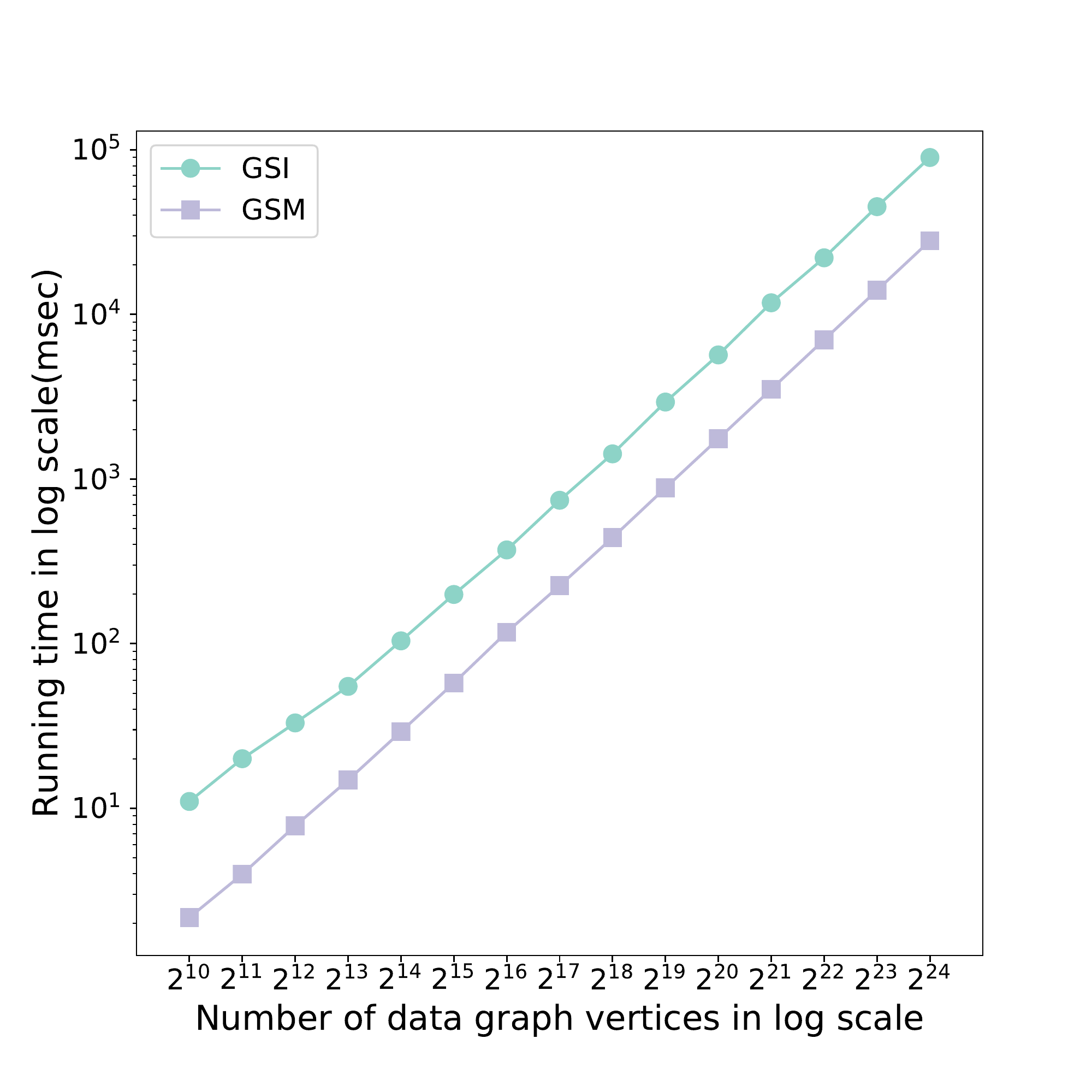}
  \end{minipage}
  \begin{minipage}{0.43\columnwidth}
    \includegraphics[width=\linewidth]{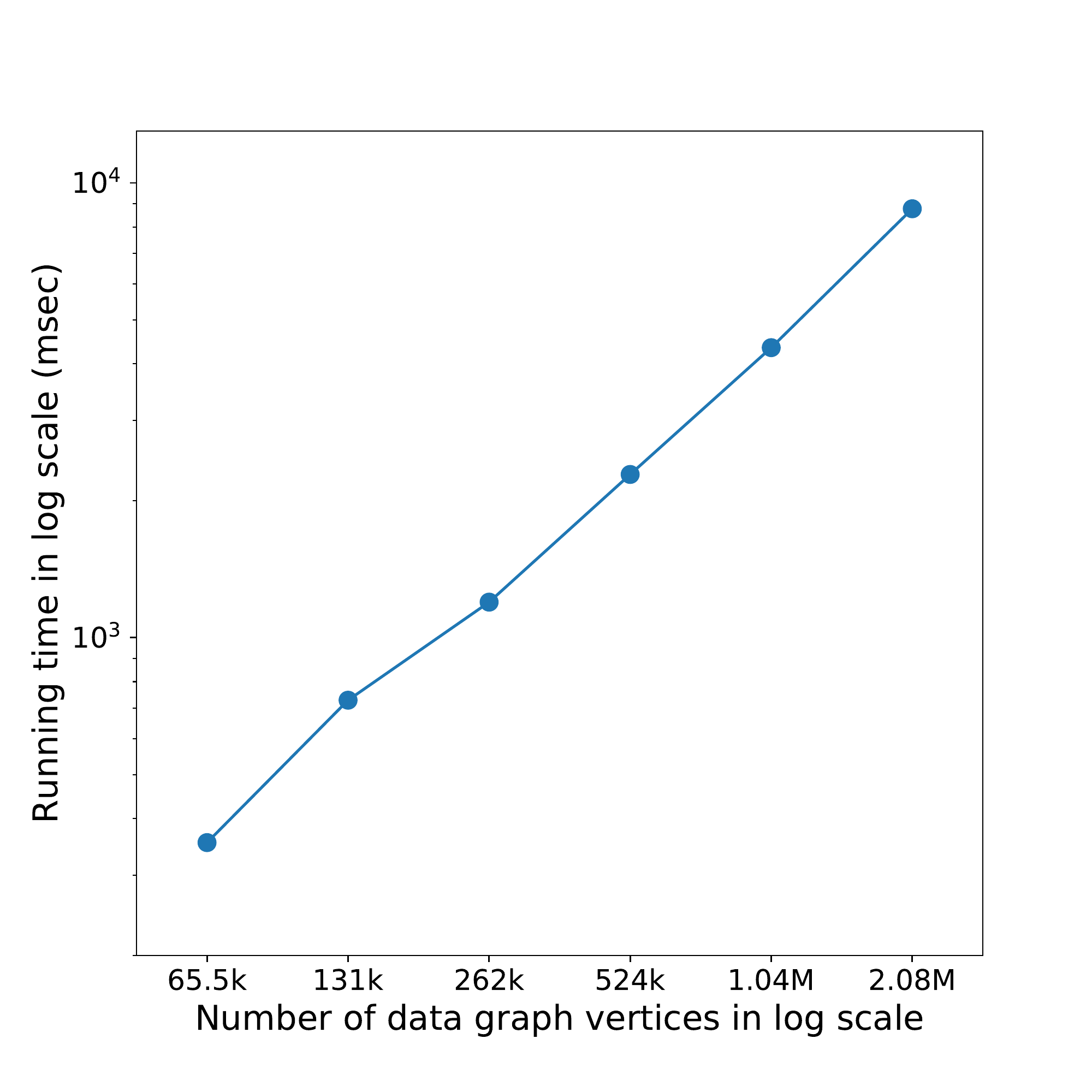}
  \end{minipage}
  \caption{Left: Scalability test of GSM and GSI on increasing sizes of the synthetic \emph{delaunay} graph ($2^{10}$--$2^{24}$ vertices). The query graph here is a triangle because GSI runs out of memory with larger queries. Right: Scalability test of GSM on increasing sizes of the \emph{kron} data graph. In this test, we used a query graph with 6 nodes and 6 edges. GSI ran out of memory on all of the \emph{kron} tests.
  }
  \label{fig:scale}
\end{figure}

\paragraph{Scalability on synthetic graphs}
We also performed scalability tests on the synthetic datasets \emph{delaunay} and \emph{kron} (Figure~\ref{fig:scale}), matching against triangles (\emph{delaunay}) and a 6-node, 6-edge graph (\emph{kron}). Our runtimes are consistently $3\times$ better than GSI on the triangle test. We also ran a larger query graph with 8 nodes and 8 edges on \emph{delaunay} datasets. On this test, GSI ran out of memory on \emph{delaunay\_n15} ($2^{15}$ nodes), while GSM successfully ran up until \emph{delaunay\_n20} ($2^{20}$ nodes).

\paragraph{Performance analysis}
In Table~\ref{tab:perf_anal}, we evaluate the optimization method of \emph{k-look-ahead} (only 1-/2-look-ahead) by removing the optimization from the implementation and calling the resulting method GSM w/o opt. We compare it with the performance of the original implementation GSM on the same set of datasets. From the table, we can see that the optimization does play a role in the acceleration on the GPU since it reduces the candidate set and thus the number of traversals for the next iteration. The optimization makes the most difference in \emph{road\_central} network. The reason is that this graph has an average node degree of 1.2, which is much smaller compared with other graphs. Our query graph has a minimum degree of 2. This graph thus contains the most number of unfruitful candidates that are pruned out by \emph{k-look-ahead}.

% \begin{table}
%   \caption{Table captions should be placed above the
%   tables.}\label{tab1}
%   \begin{tabular}{|l|l|l|}
      %       \hline
      %       Heading level &  Example & Font size and style\\
      %       \hline
      %       Title (centered) &  {\Large\bfseries Lecture Notes} & 14 %point, bold\\
      %       1st-level heading &  {\large\bfseries 1 Introduction} & %12 point, bold\\
      %       2nd-level heading & {\bfseries 2.1 Printing Area} & 10 %point, bold\\
      %       3rd-level heading & {\bfseries Run-in Heading in Bold.} %Text follows & 10 point, bold\\
      %       4th-level heading & {\itshape Lowest Level Heading.} %Text follows & 10 point, italic\\
      %       \hline
      %     \end{tabular}
      %       \end{table}

\begin{table}
  \caption{Runtime comparison of GSM with (column 2) and without (column 3) the k-look-ahead optimization.\label{tab:perf_anal}}
  \centering
  \setlength{\tabcolsep}{3pt}
  \begin{tabular}{llll} \toprule Dataset & GSM (ms) & GSM w/o opt (ms) & Speedup\\
    \midrule
    enron & 290.14 & 298.84 & 1.03
    \\ gowalla & 1690.93 & 1775.47 & 1.05
    \\ patent & 3976.66 & 4095.96 & 1.03
    \\ road\_central & 1318.83 & 1819.99 & 1.38
    \\ \bottomrule
  \end{tabular}

\end{table}

%% file: tex/conc.tex
In this paper, we introduce an efficient and scalable method, GSM, that effectively leverages the massive parallel resources of the GPU\@. In the process of developing GSM, we grew to appreciate the importance of memory management, a vital technique when trying to solve a problem that is naturally expressed as a depth-first traversal when we chose to use a BFS-based approach. Inefficient memory management would have led to the generation of a large amount of intermediate results and exhausted the limited GPU memory quickly. We expect that future fruitful work will both look at matching larger subgraphs (larger than the up-to-20 element graphs we study in this work) and methods that further prioritize efficient memory usage.